\documentclass[epj]{svjour}

\usepackage{here}
\usepackage{graphicx}
\usepackage{subfigure}
\begin{document}

\title{The Fourth Standard Model Family and the Competition in Standard Model Higgs Boson Search at Tevatron and LHC}
\titlerunning{The $4^{th}$ SM Family and the Competition in SM Higgs Search at Tevatron and LHC}

\author{N. Becerici Schmidt\inst{1}, S. A. \c Cetin\inst{2}, S. I{\c s}t{\i}n \inst{1} \and S. 
Sultansoy\inst{3,4}}
\authorrunning{N. Becerici Schmidt \textit{et al}}

\institute{
Physics Department, Bo\u gazi{\c c}i University, Istanbul, Turkey
 \and
Physics Division, Do\u gu{\c s}  University, Istanbul, Turkey
 \and
Physics Division, TOBB University of Economics and Technology, Ankara, Turkey
 \and
Institute of Physics, Academy of Sciences, Baku, Azerbaijan
}

%

\abstract{The impact of the fourth Standard Model family on Higgs boson search at Tevatron and LHC is 
reviewed.}


\maketitle

\section{Introduction}
\label{intro}
Recent changes in the schedule of the LHC operation has resulted in an additional two years extension of Tevatron discovery 
challenges in the search for the Higgs boson (H), the fourth Standard Model family and so on (including SUSY). The fourth family is 
a natural consequence of the Standard Model (SM) basic principles and the actual patterns of the first three family fermion masses 
and mixings \cite{fritzsch,datta,celikel} (for reviews see \cite{sultansoy1,sultansoy2,frampton,sultansoy3,sultansoy4,sultansoy5,holdom}). 
We should once again note that, in contrast to the widespread opinion, electroweak precision data does not exclude the fourth 
family\cite{he,novikov,kribs,ozcan}. The fourth family matters were discussed in detail during the topical workshop held in September 
2008 at CERN 
\cite{b3sm} (see \cite{holdom} for resume of the workshop). 

Concerning the Higgs boson, the existence of the fourth Standard Model family has a strong impact on the search 
strategies of Tevatron and LHC \cite{kribs,arik1,ginzburg,sultansoy6,arik2,cakir1,arik3,arik4,arik5} mainly due 
to the essential enhancement of the gg$\rightarrow$H production channel. In this paper, SM-4 (SM-3) denotes Standard Model with 4 
(3) families. 

\section{Fourth SM family effects on the Higgs boson}
\label{sec1}
\begin{figure}[H]
\begin{center}
\resizebox*{7.45cm}{!}{
\includegraphics[width=0.5\textwidth]{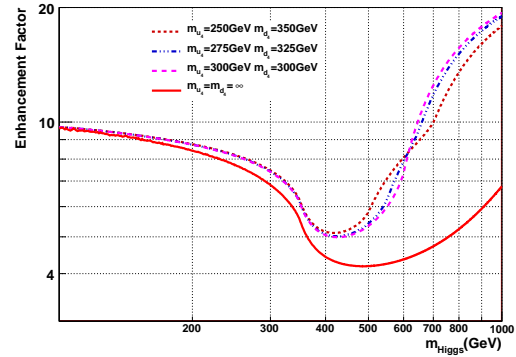}}
\caption{Enhancement factors for the SM Higgs production via gluon fusion 
when the fourth SM family quark masses are around 300GeV. Enhancement factors in the infinite mass limit are also shown for 
comparison.}
\label{fig:enh1}
\end{center}
\end{figure}

\begin{figure}[H]
\begin{center}
\resizebox*{7.45cm}{!}{
\includegraphics[width=0.5\textwidth]{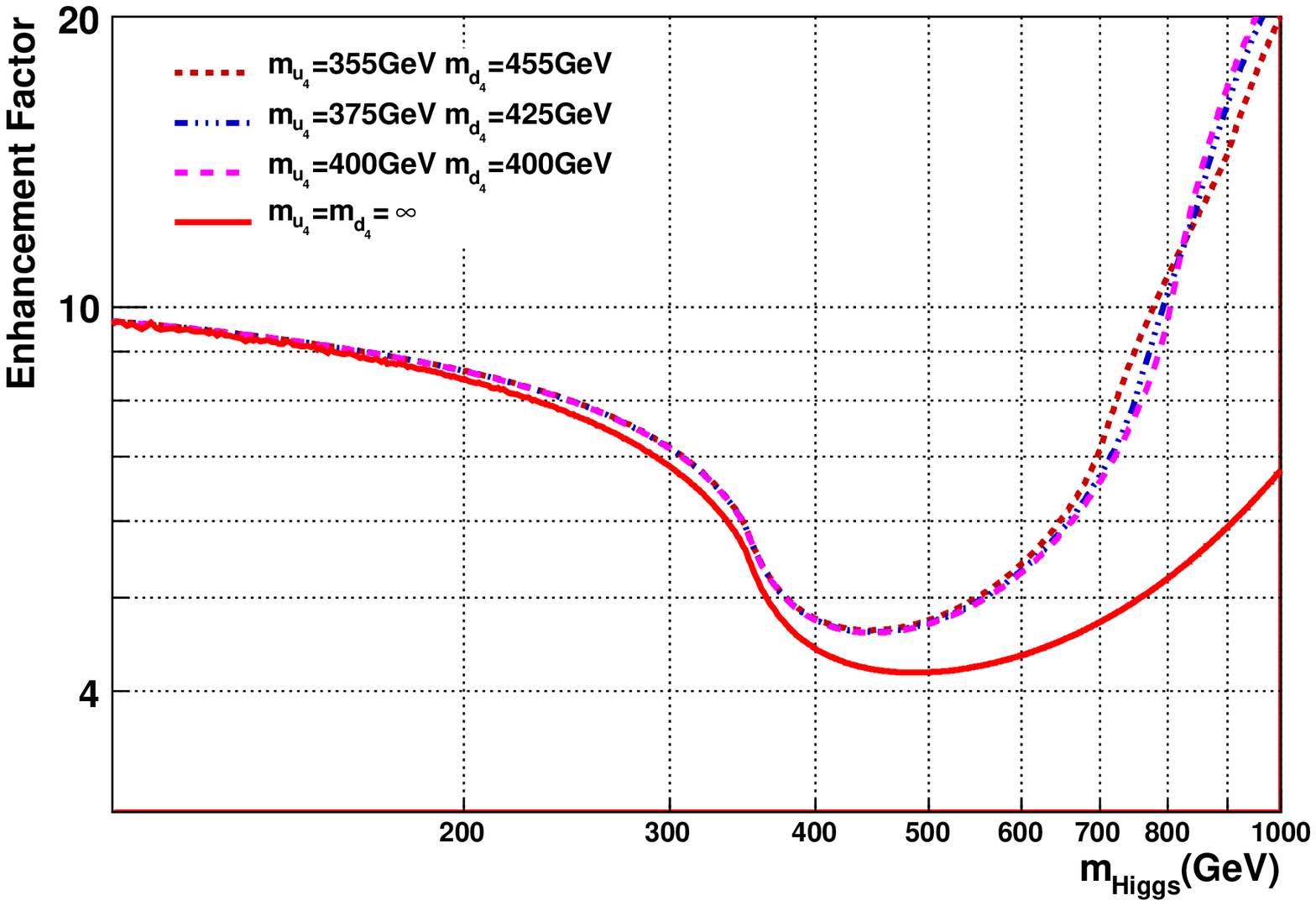}}
\caption{The same as Figure \ref{fig:enh1} but for fourth SM family quark masses around 400GeV.}
\label{fig:enh2}
\end{center}
\end{figure}

\begin{figure}[H]
\begin{center}
\resizebox*{7.7cm}{!}{
\includegraphics[width=0.5\textwidth]{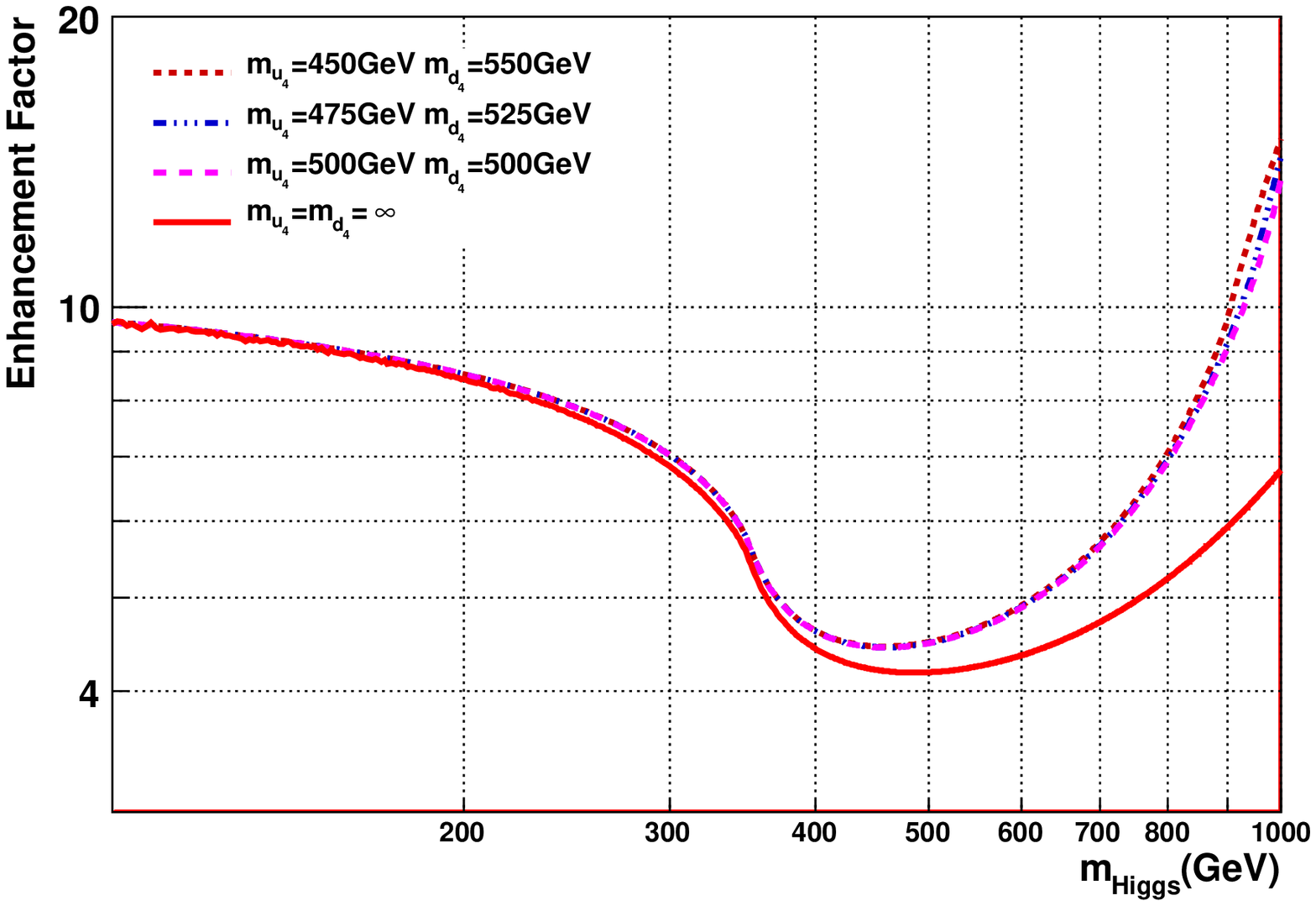}}
\caption{The same as Figure \ref{fig:enh1} but for fourth SM family quark masses around 500GeV.}
\label{fig:enh3}
\end{center}
\end{figure}

\begin{figure}[H]
\begin{center}
\resizebox*{7.7cm}{!}{
\includegraphics[width=0.5\textwidth]{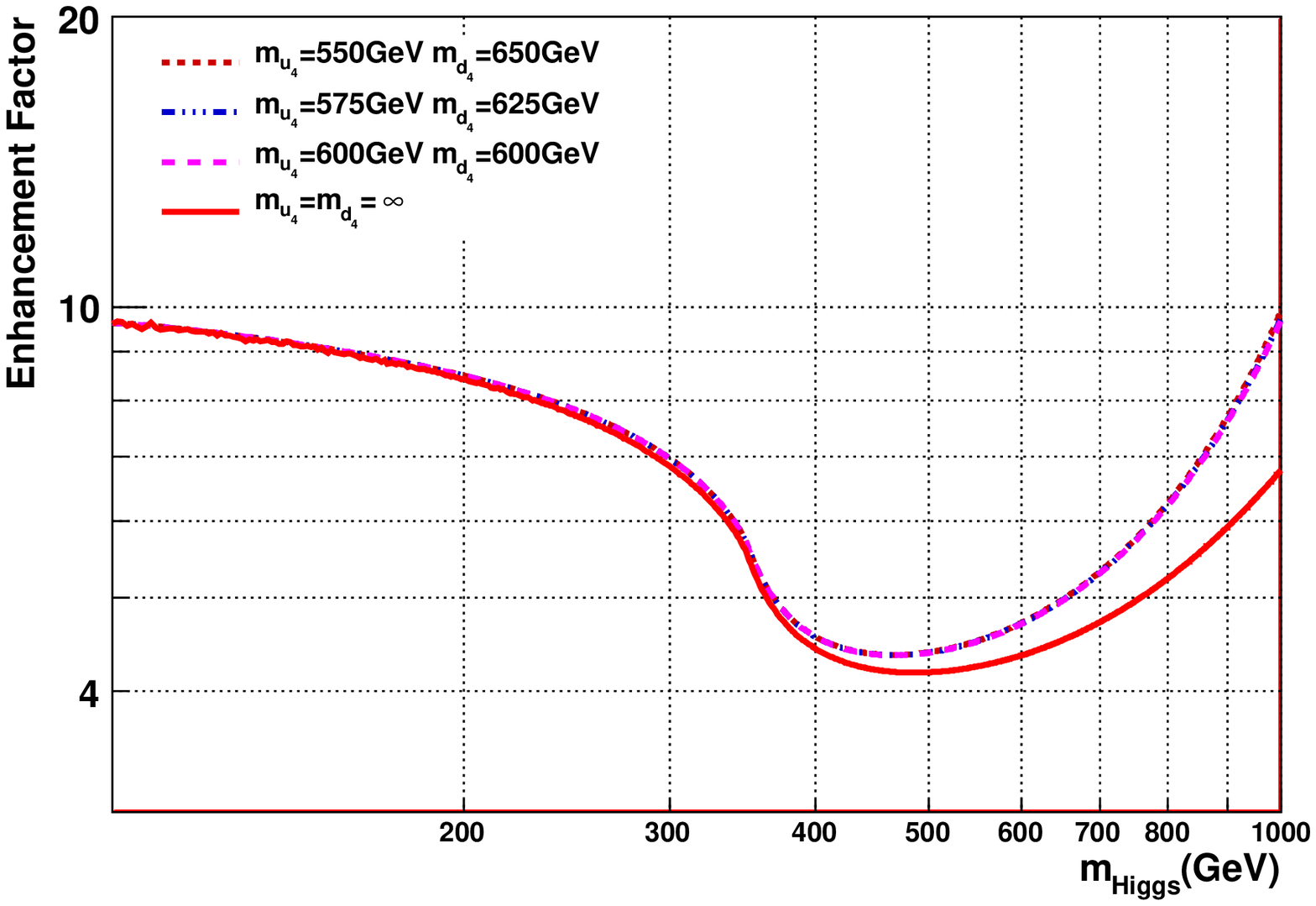}}
\caption{The same as Figure \ref{fig:enh1} but for fourth SM family quark masses around 600GeV.}
\label{fig:enh4}
\end{center}
\end{figure}

\begin{figure}[H]
\begin{center}
\resizebox*{7.7cm}{!}{
\includegraphics[width=0.5\textwidth]{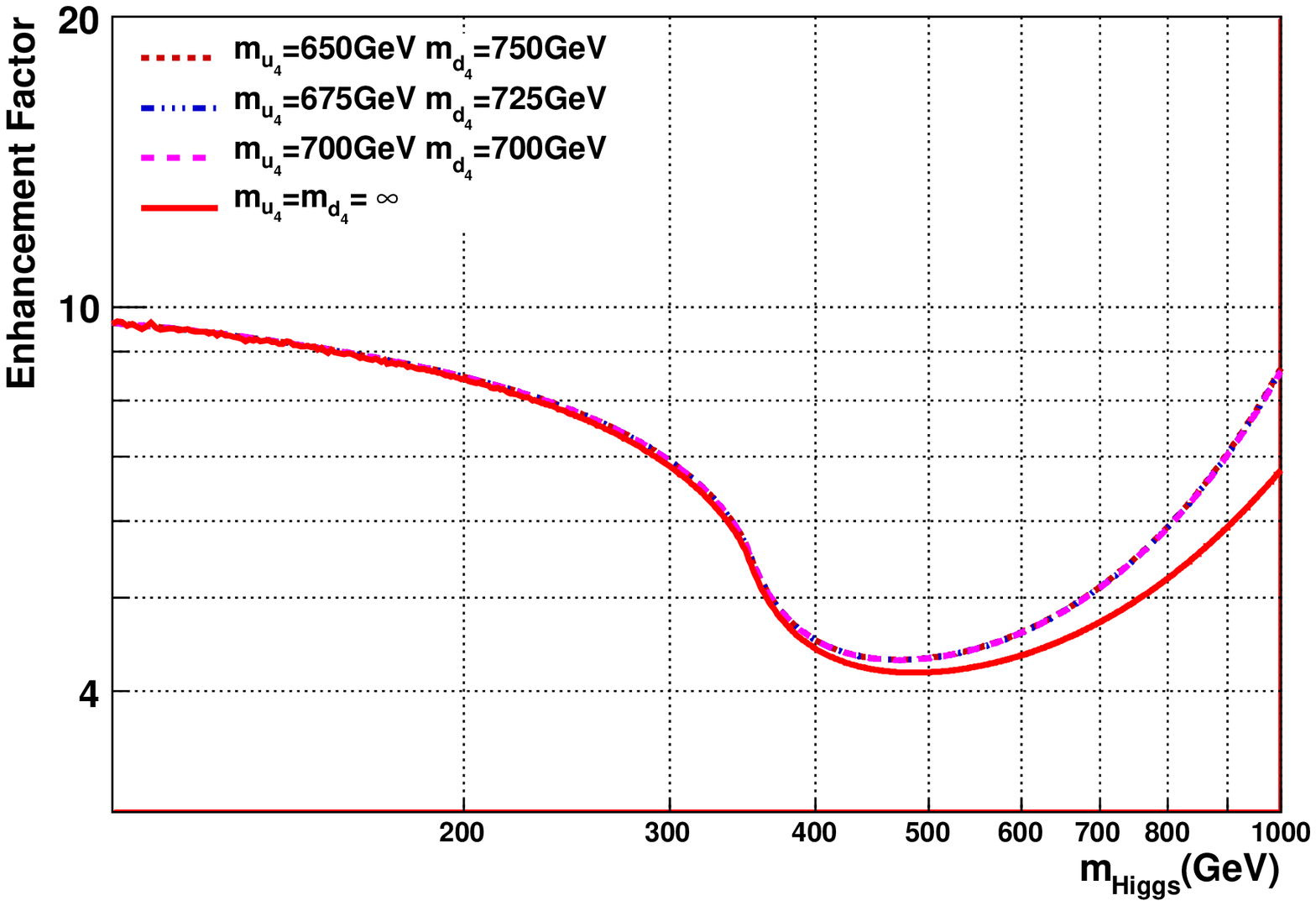}}
\caption{The same as Figure \ref{fig:enh1} but for fourth SM family quark masses around 700GeV.}
\label{fig:enh5}
\end{center}
\end{figure}

The crucial contribution of the new heavy quarks to the gg$\rightarrow$H vertex via the triangular loop has been realized many years 
ago \cite{barger}. Additional quark loops introduced by the fourth SM family quarks strengthens the gg$\rightarrow$H vertex by a 
factor of about 3, hence causing an enhancement in the cross section by about 9. The actual values depend on 
the mass of the Higgs boson and fourth SM family fermions. Figures \ref{fig:enh1}-\ref{fig:enh5} demonstrate this 
dependence for different scenarios. As seen from these figures, the choice of infinitely heavy fourth 
SM family quarks corresponds to the 
most conservative scenario which will be the assumption in the rest of this work.


\newpage

\begin{figure} [H]
\begin{center}
\resizebox*{8.5cm}{!}{
\includegraphics[width=0.3\textwidth]{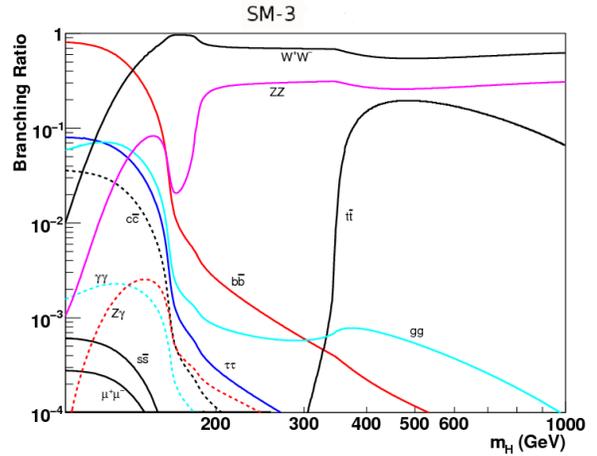}}
\caption{Higgs branching ratios in SM-3.}
\label{fig:br3}
\end{center}
\end{figure}

\begin{figure} [H]
\begin{center}
\resizebox*{8.5cm}{!}{
\includegraphics[width=0.3\textwidth]{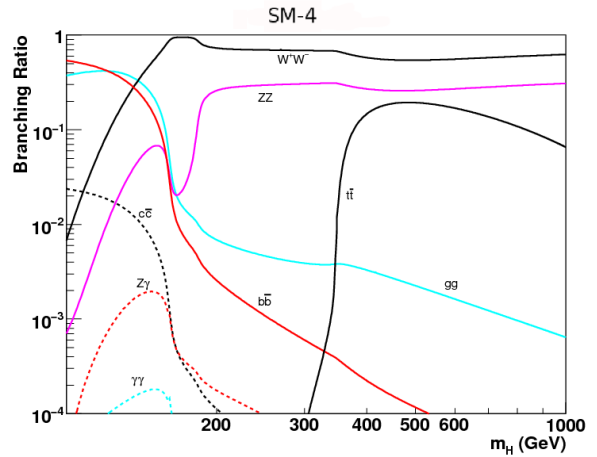}}
\caption{Higgs branching ratios in SM-4 where fourth family fermions are assumed to be infinitely heavy.}
\label{fig:br4}
\end{center}
\end{figure}

The fourth SM family fermions will affect a number of other vertices 
\cite{kribs,arik1,ginzburg,sultansoy6,arik2,cakir1,arik3,arik4,arik5,gokhan1,gokhan2} along with gg$\rightarrow$H resulting 
in new branching ratio values of the Higgs decays. 
Figure \ref{fig:br3} illustrates Higgs decay branching ratios in SM-3 and Figure \ref{fig:br4} in SM-4 with 
infinitely heavy fourth family. 

In principle if the neutrino has Majorana nature, $\nu_{4}$ could be essentialy lighter than the other members of the fourth family 
and the Higgs boson could decay into the fourth family neutrinos; 
this scenario is considered in \cite{gokhan1,gokhan2}.

\newpage  
\clearpage

\section{The Tevatron Perspective}
\label{sec:2}
\begin{figure}[H]
\begin{center}
\resizebox*{8.cm}{!}{
\includegraphics[width=0.5\textwidth]{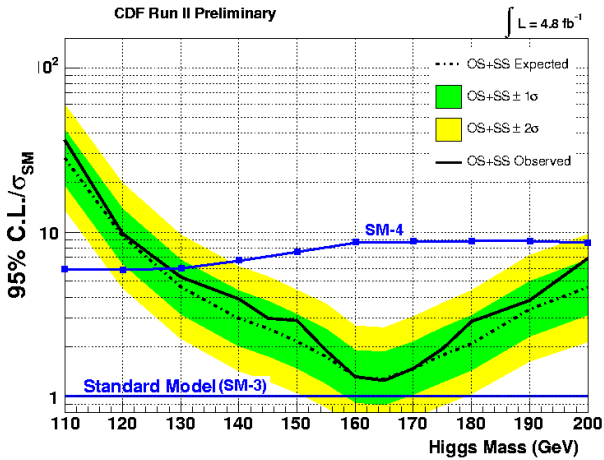}}
\caption{Exclusion plot from CDF \cite{fnal-cdf} experiment.}
\label{fig:fnal-cdf}
\end{center}
\end{figure}

\begin{figure}[H]
\begin{center}
\resizebox*{8.cm}{!}{
\includegraphics[width=0.5\textwidth]{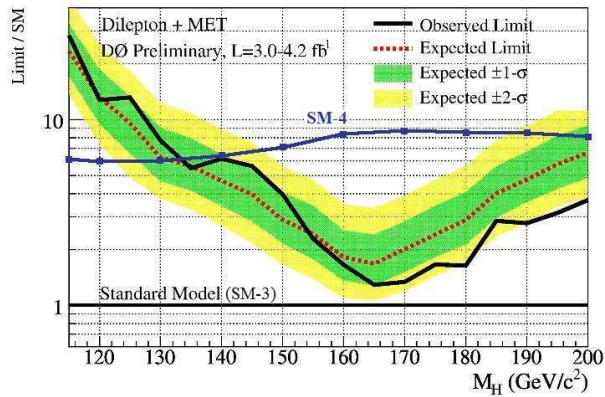}}
\caption{Exclusion plot from D0 \cite{fnal-d0} experiment.}
\label{fig:fnal-d0}
\end{center}
\end{figure}

$gg\rightarrow H\rightarrow WW\rightarrow \ell \ell \nu\nu$ (where $\ell$ denotes $e$ or $\mu$) is the most promising channel in SM-4 
case. Figures \ref{fig:fnal-cdf} and \ref{fig:fnal-d0} show recent results 
\cite{fnal-cdf,fnal-d0} on this channel where we add the curves corresponding to SM-4. It is clear that Higgs boson with mass 
130 - 200 GeV is excluded if a fourth SM family exists while only the 160 - 170 GeV 
region is excluded in the SM-3 case. As seen from Figure \ref{fig:fnal-d0}, D0 actually excludes even higher Higgs masses 
(presumably up to 240 GeV) in SM-4, however the analysis ends at 200 GeV.

Although the contribution coming from WH, ZH and VBF processes to the total production cross section is about 20$\%$, the selection 
criteria for $H\rightarrow WW $ signature would suppress this contribution to the level of a few percent. Hence, the current CDF and 
D0 results shown in Figures \ref{fig:fnal-cdf} and \ref{fig:fnal-d0} can be considered to be coming from $gg\rightarrow H$ alone. 

Taking into account the fact that nature could prefer the SM-4 case, both D0 and CDF 
should extend the horizontal axis up to 300 GeV and, moreover, combine 
their results on the $WW$ channel.
Furthermore, combined analysis of all channels and both experiments done for SM-3 should be repeated for SM-4. 
Examples of proper approach are 
\cite{tevatron1,tevatron2,tevatron3,tevatron4,tevatron5,tevatron6,tevatron7,tevatron8,tevatron9,tevatron10}.


\section{The LHC Perspective}
\label{sec:2}
As an example for the LHC perspectives, we restrict ourselves to a detailed consideration of the \textit{Golden Mode} at the ATLAS 
experiment \cite{atlas-jinst,atlas-csc}. A similar analysis can be carried out for CMS 
as well. Moreover,  a combined analysis of both LHC experiments could be useful.

\begin{figure}[H]
\begin{center}
\resizebox*{8.8cm}{!}{
\subfigure[]{
\includegraphics[width=0.5\textwidth]{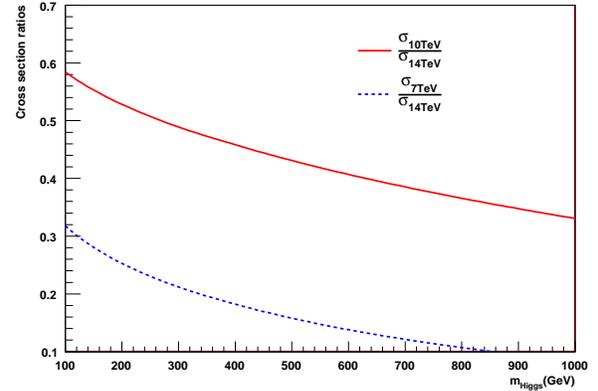}}
}
\resizebox*{8.8cm}{!}{
\subfigure[]{
\includegraphics[width=0.5\textwidth]{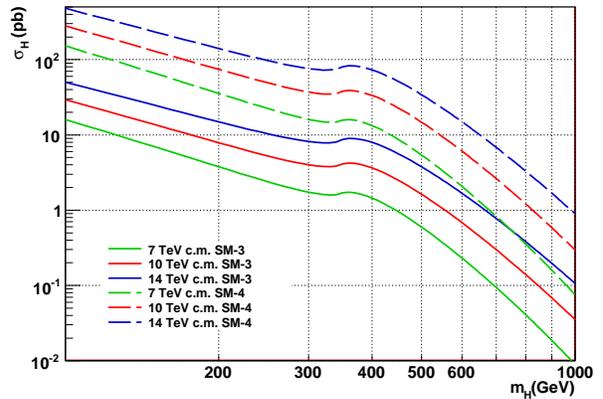}}
}
\caption{a)Ratios of the Higgs production cross sections via gluon fusion at 10 and 7 TeV center of
mass energies to 14 TeV center of mass energy; b)Higgs production cross section via 
gluon fusion at different center of mass energies for SM-3 and SM-4 cases.}
\label{fig:xscn}
\end{center}
\end{figure}

The design center of mass energy of 14 TeV is the basic scenario, in addition, we also 
consider 10 TeV and 7 TeV cases for early phase operation. As input parameters, we use the most recent ATLAS 
simulation 
results for 14 TeV published in \cite{atlas-csc}. The analysis for 10 and 7 TeV cases is performed using the Higgs 
production cross section ratio given in Figure \ref{fig:xscn}a (calculations are performed using HIGLU 
\cite{higlu}). The backgrounds considered in \cite{atlas-csc} are rescaled using the calculations performed in COMPHEP \cite{comphep} 
in a similar manner. It is recently shown in \cite{bruce} that the theoretical uncertainties on the SM background via two weak boson 
production is around 5-20$\%$. This uncertainity merely effects the significance results presented in this section.


\subsection{$\sqrt{s}=14$ TeV case}
\label{subsec:14}
\begin{figure}[H]
\begin{center}
\centering
\resizebox{9.cm}{8.cm}{
\includegraphics[width=0.5\textwidth]{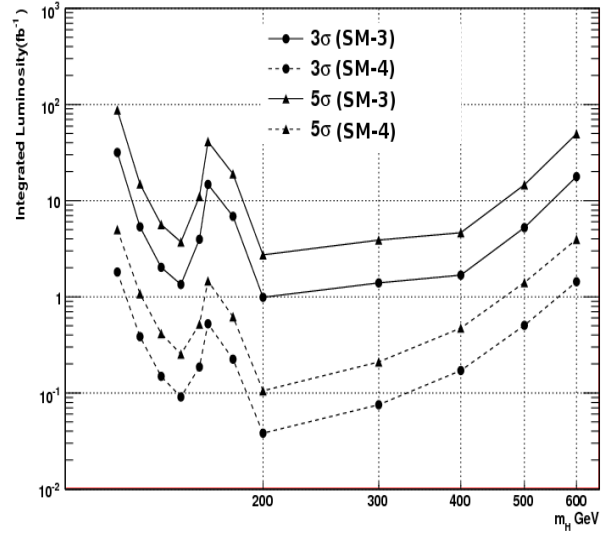}}
\caption{Integrated luminosity needed at 14 TeV for 3$\sigma$ and 5$\sigma$ for $gg \rightarrow H \rightarrow 
ZZ^{(*)}\rightarrow4\ell$ channel considering SM-3 and SM-4 cases.}
\label{fig:int14}
\end{center}
\end{figure}

The ATLAS simulation results for the $gg\rightarrow H \rightarrow ZZ^{(*)}\rightarrow4\ell$ 
signature are presented in column 2 and 4 of Table \ref{tab:xscn14} where we add the SM-4 case in column 3. 
Using these foreseen reconstructed signal and background cross sections and the statistical 
significance (SS) formula
\begin{center}
$\sqrt{2(s+b)ln(1+s/b)-2s}$ \\ 
\end{center}
we calculate SS for different integrated luminosities 
($L_{int}$) as shown in Table \ref{tab:ss14}. The necessary $L_{int}$ values to achieve $3\sigma$ and $5\sigma$ 
significance are also shown in Table \ref{tab:int14} and plotted in Figure \ref{fig:int14}. 

\begin{table}[H]
\caption{Expected cross sections of the reconstructed $gg\rightarrow H\rightarrow ZZ^{(*)}$$\rightarrow4\ell$ 
channel and its total background at 14 TeV.}
\label{tab:xscn14}
\tiny
\centering
\begin{tabular}{|c||c|c|c|}
\hline
\multicolumn{4}{|c|}{ $gg\rightarrow H \rightarrow ZZ^{(*)}$$\rightarrow4\ell$ at 14 TeV}  \\
\hline
$ \rm {m_H}$ (GeV)& \multicolumn{3}{|c|}{ Cross section (fb) }           \\
\hline
                & \multicolumn{2}{|c|}{\bf Signal}                      & \multicolumn{1}{|c|}{\bf Background} \\
                & SM-3                 & SM-4   & SM-3 \& SM-4  \\
\hline
 120            & 0.281                & 1.658  & 0.198         \\
 130            & 0.816                & 4.902  & 0.197         \\
 140            & 1.511                & 9.885  & 0.189         \\
 150            & 1.94                 & 14.214 & 0.172         \\
 160            & 1.03                 & 8.692  & 0.223         \\
 165            & 0.484                & 4.197  & 0.253         \\
 180            & 1.32                 & 11.339 & 0.951         \\
 200            & 6.68                 & 55.935 & 3.09          \\
 300            & 4.21                 & 28.697 & 1.65          \\
 400            & 3.34                 & 14.747 & 1.21          \\
 500            & 1.66                 & 6.937  & 1.14          \\
 600            & 0.76                 & 3.308  & 0.914         \\
\hline
\end{tabular}
\end{table}

\begin{table}[H]
\caption{Expected statistical significance for various integrated luminosity values at 14 TeV.}
\label{tab:ss14}
\tiny
\centering
\begin{tabular}{|c|c|c|c|c|c|c|c|c|}
\hline
\multicolumn{9}{|c|}{ $gg\rightarrow H \rightarrow ZZ^{(*)}$$\rightarrow4\ell$ at 14 TeV}  \\
\hline
m$_H$                        & \multicolumn{2}{|c|}{0.3fb$^{-1}$} & \multicolumn{2}{|c|}{ 1fb$^{-1}$ }& 
\multicolumn{2}{|c|}{ 3fb$^{-1}$}
& \multicolumn{2}{|c|}{ 10fb$^{-1}$} \\
 (GeV)               & SM-3 & SM-4           & SM-3 & SM-4           & SM-3 & SM-4           & SM-3 & SM-4 \\
\hline
 120            & 0.29 & 1.22                           & 0.53 & 2.23           & 0.92 & 3.87           & 1.68 & 7,07 \\
 130            & 0.71 & 2.65           & 1.29 & 4.83           & 2.24 & 8.37           & 4.10 & 15,29 \\
 140            & 1.15 & 4.25           & 2.10 & 7.77           & 3.65 & 13.45          & 6.66 & 24,56 \\
 150            & 1.42 & 5.45           & 2.59 & 9.95           & 4.48 & 17.23          & 8.19 & 31,45 \\
 160            & 0.82 & 3.81           & 1.50 & 6.95           & 2.60 & 12.05          & 4.76 & 21,99  \\
 165            & 0.42 & 2.27           & 0.78 & 4.14           & 1.35 & 7.17           & 2.46 & 13,08 \\
 180            & 0.62 & 3.47           & 1.14 & 6.34           & 1.98 & 10.98           & 3.62 & 20,06 \\
 200            & 1.65 & 8.42           & 3.02 & 15.37          & 5.23 & 26.63          & 9.55 & 48,61 \\
 300            & 1.39 & 5.98           & 2.53 & 10.92           & 4.39 & 18.92          & 8.02 & 34,55 \\
 400            & 1.27 & 3.98           & 2.31 & 7.27           & 4.01 & 12.59          & 7.33 & 22,98 \\
 500            & 0.71 & 2.31           & 1.30 & 4.21           & 2.26 & 7.30           & 4.13 & 13,32 \\
 600            & 0.39 & 1.15           & 0.71 & 2.51           & 1.23 & 4.35           & 2.25 & 7,94 \\
\hline
\end{tabular}
\end{table}

\begin{table}[H] 
\caption{Integrated luminosity needed for 3$\sigma$ and 5$\sigma$ significance at 14 TeV.}
\label{tab:int14}
\tiny
\centering
\begin{tabular}{|c||c|c|c|c|}
\hline
\multicolumn{5}{|c|}{ $gg \rightarrow H \rightarrow ZZ^{(*)}\rightarrow4\ell$ at 14 TeV} \\
\hline
$ \rm {m_H}$ (GeV)      & \multicolumn{4}{|c|}{ Luminosity (fb$^{-1}$)} \\
\hline
                        & \multicolumn{2}{|c|}{\bf for 3$\sigma$} & \multicolumn{2}{|c|}{\bf for 5$\sigma$} \\
                        & SM-3          & SM-4         & SM-3     & SM-4          \\
                        & (14 TeV)      & (14 TeV)     & (14 TeV) & (14 TeV)      \\
\hline
120                     & 31.65         & 1.80         & 87.92    & 5.00          \\
130                     & 5.34          & 0.38         & 14.83    & 1.06          \\
140                     & 2.02          & 0.15         & 5.62     & 0.41          \\
150                     & 1.34          & 0.01         & 3.72     & 0.25          \\
160                     & 3.97          & 0.19         & 11.03    & 0.52          \\
165                     & 14.80         & 0.52         & 41.11    & 1.46          \\
180                     & 6.85          & 0.22         & 19.03    & 0.62          \\
200                     & 0.98          & 0.04         & 2.73     & 0.10          \\
300                     & 1.39          & 0.07         & 3.88     & 0.21          \\
400                     & 1.67          & 0.17         & 4.65     & 0.47          \\
500                     & 5.25          & 0.51         & 14.60    & 1.41          \\
600                     & 17.78         & 1.42         & 49.40    & 3.97          \\
\hline
\end{tabular}
\end{table}

It is clear that with only 500 
pb$^{-1}$ Higgs boson will be observed at $3\sigma$ level in the golden mode for the SM-4 case if the mass of the Higgs is 
between 130 - 500 GeV. An integrated luminosity of 100 pb$^{-1}$ will be more than enough to scan 200 - 
300 GeV Higgs at $3\sigma$ level. One should note that 130 - 200 GeV Higgs in SM-4 is already excluded by 
Tevatron.


\subsection{$\sqrt{s}=10$ TeV case}
\label{subsec:10}
\begin{figure}[H]
\begin{center}
\centering
\resizebox{9.cm}{8.cm}{
\includegraphics[width=0.5\textwidth]{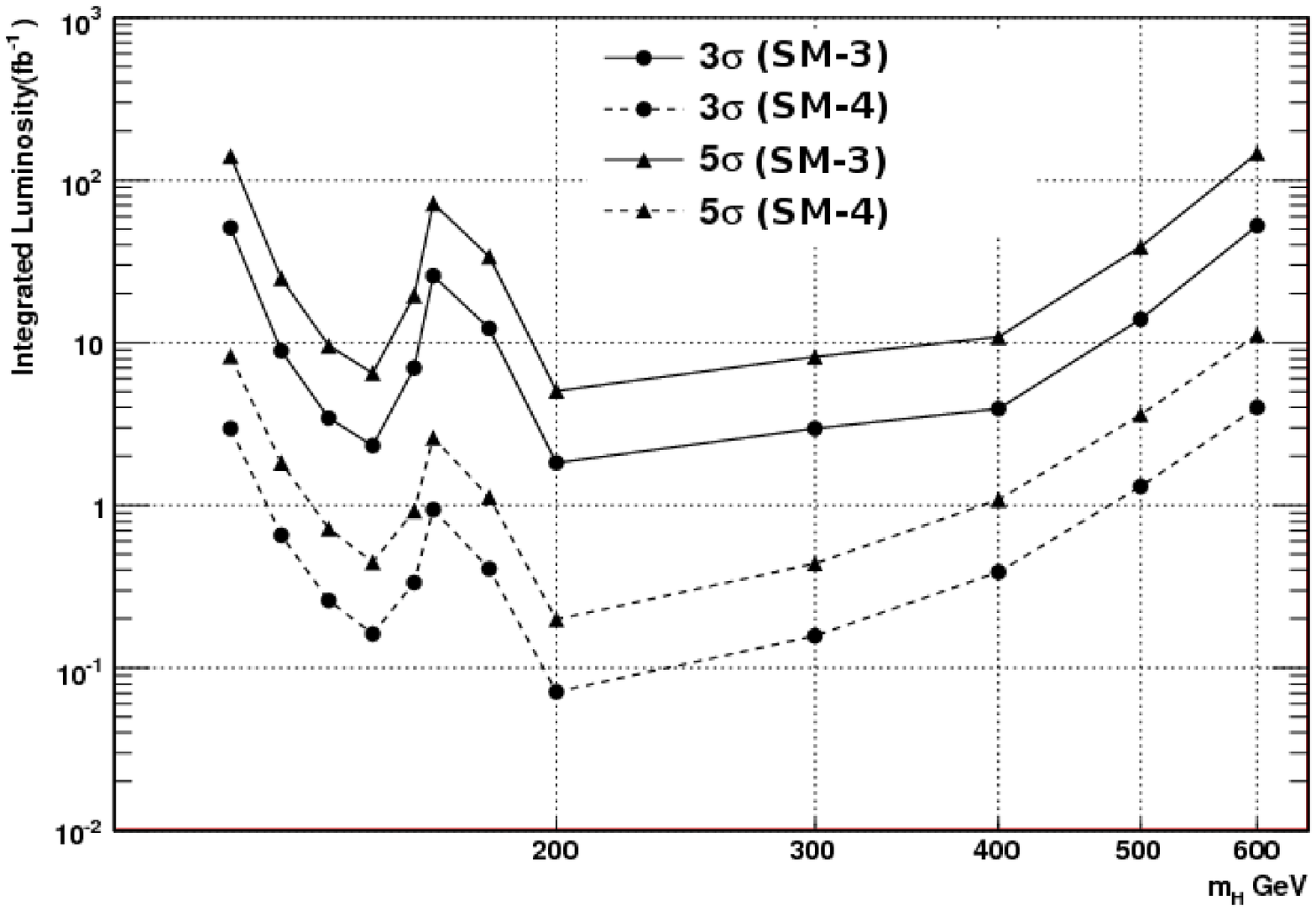}}
\caption{Integrated luminosity needed at 10 TeV for 3$\sigma$ and 5$\sigma$ for $gg \rightarrow H \rightarrow 
ZZ^{(*)}\rightarrow4\ell$ channel considering SM-3 and SM-4 cases.}
\label{fig:int10}
\end{center}
\end{figure}

Table \ref{tab:xscn10} shows expected cross sections after reconstruction of $gg\rightarrow H \rightarrow 
ZZ^{(*)}\rightarrow4\ell$ channel and its backgrounds. Corresponding statistical significance for 
various integrated luminosities are shown in Table \ref{tab:ss10} and $L_{int}$ needed for $3\sigma$ and $5\sigma$ 
significance are given in Table \ref{tab:int10} and plotted in Figure \ref{fig:int10}. 

\begin{table}[H]
\caption{Expected cross sections of the reconstructed $gg\rightarrow H\rightarrow ZZ^{(*)}$$\rightarrow4\ell$
channel and its total background at 10 TeV.}
\label{tab:xscn10}
\tiny
\centering
\begin{tabular}{|c||c|c|c|}
\hline
         \multicolumn{4}{|c|}{ $gg\rightarrow H \rightarrow ZZ^{(*)}$$\rightarrow4\ell$ at 10 TeV}  \\
\hline
$ \rm {m_H}$ (GeV)& \multicolumn{3}{|c|}{ Cross section (fb) }           \\
\hline
                & \multicolumn{2}{|c|}{\bf Signal}                      & \multicolumn{1}{|c|}{\bf Background} \\
                & SM-3          & SM-4        & SM-3 \& SM-4   \\
                &(10 TeV)       &(10 TeV)     & (10 TeV)       \\
\hline
 120            & 0.160         & 0,948       & 0.100  \\
 130            & 0.460         & 2,763       & 0.100  \\
 140            & 0.844         & 5,521       & 0.096  \\
 150            & 1.07          & 7,840       & 0.087  \\
 160            & 0.56          & 4,726       & 0.113  \\
 165            & 0.263         & 2,280       & 0.128  \\
 180            & 0.71          & 6,099       & 0.484  \\
 200            & 3.52          & 29,475      & 1.572  \\
 300            & 2.05          & 13,974      & 0.839  \\
 400            & 1.53          & 6,755       & 0.615  \\
 500            & 0.71          & 2,967       & 0.580  \\
 600            & 0.31          & 1,349       & 0.465  \\
\hline
\end{tabular}
\end{table}

\begin{table}[H]
\caption{Expected statistical significance for various integrated luminosity values at 10 TeV.}
\label{tab:ss10}
\tiny
\centering
\begin{tabular}{|c|c|c|c|c|c|c|c|c|}
\hline
\multicolumn{9}{|c|}{ $gg\rightarrow H \rightarrow ZZ^{(*)}$$\rightarrow4\ell$ at 10 TeV}  \\
\hline
m$_H$                        & \multicolumn{2}{|c|}{0.1fb$^{-1}$ } & \multicolumn{2}{|c|}{0.2fb$^{-1}$ } & 
\multicolumn{2}{|c|}{0.3fb$^{-1}$ }
& \multicolumn{2}{|c|}{0.5fb$^{-1}$ } \\
   (GeV)             & SM-3 & SM-4           & SM-3 & SM-4           & SM-3 & SM-4           & SM-3 & SM-4 \\
\hline
 120            & 0.13 & 0.55           & 0.19 & 0.78           & 0.23 & 0.95           & 0.30 & 1.23 \\
 130            & 0.31 & 1.17           & 0.45 & 1.65           & 0.55 & 2.02           & 0.71 & 2.61 \\
 140            & 0.51 & 1.86           & 0.72 & 2.63           & 0.88 & 3.22           & 1.14 & 4.16 \\
 150            & 0.62 & 2.36           & 0.88 & 3.34           & 1.07 & 4.09           & 1.39 & 5.28 \\
 160            & 0.35 & 1.64           & 0.50 & 2.32           & 0.62 & 2.84           & 0.80 & 3.67 \\
 165            & 0.18 & 0.98           & 0.26 & 1.38           & 0.32 & 1.69           & 0.42 & 2.18 \\
 180            & 0.27 & 1.49           & 0.38 & 2.10           & 0.47 & 2.57           & 0.61 & 3.32 \\
 200            & 0.70 & 3.55           & 0.99 & 5.02           & 1.21 & 6.15           & 1.57 & 7.94 \\
 300            & 0.55 & 2.39           & 0.78 & 3.37           & 0.95 & 4.13           & 1.23 & 5.34 \\
 400            & 0.48 & 1.52           & 0.68 & 2.15           & 0.83 & 2.63           & 1.07 & 3.39 \\
 500            & 0.25 & 0.83           & 0.36 & 1.17           & 0.44 & 1.43           & 0.57 & 1.85 \\
 600            & 0.13 & 0.47           & 0.18 & 0.67           & 0.23 & 0.81           & 0.29 & 1.05 \\
\hline
\end{tabular}
\end{table}

It is seen that 200 - 250 GeV Higgs will 
be covered by 100 pb$^{-1}$ and an additional 100 pb$^{-1}$ will increase the reach up to 350 GeV.

\begin{table}[H]
\caption{Integrated luminosity needed for 3$\sigma$ and 5$\sigma$ significance at 10 TeV.}
\label{tab:int10}
\label{CrossSections_3_4}
\tiny
\centering
\begin{tabular}{|c||c|c|c|c|}
\hline
\multicolumn{5}{|c|}{ $gg\rightarrow H \rightarrow ZZ^{(*)}\rightarrow4\ell$ at 10 TeV} \\
\hline
$ \rm {m_H}$ (GeV)      & \multicolumn{4}{|c|}{ Luminosity (fb$^{-1}$)} \\
\hline
                        & \multicolumn{2}{|c|}{\bf for 3$\sigma$}               & \multicolumn{2}{|c|}{\bf for 5$\sigma$} \\
                        & SM-3   & SM-4          & SM-3       & SM-4\\
                        &(10TeV) & (10 TeV)      & (10 TeV)   & (10 TeV)\\
\hline
120                     & 50.88  & 2.97          & 141.35     & 8.26 \\
130                     & 8.91   & 0.65          & 24.76      & 1.83  \\
140                     & 3.46   & 0.26          & 9.61       & 0.72  \\
150                     & 2.33   & 0.16          & 6.49       & 0.45  \\
160                     & 7.02   & 0.33          & 19.50      & 0.93  \\
165                     & 25.91  & 0.94          & 72.00      & 2.61  \\
180                     & 12.22  & 0.40          & 33.95      & 1.13  \\
200                     & 1.82   & 0.07          & 5.07       & 0.20  \\
300                     & 2.95   & 0.16          & 8.21       & 0.44  \\
400                     & 3.91   & 0.38          & 10.87      & 1.08  \\
500                     & 14.01  & 1.30          & 38.91      & 3.62  \\
600                     & 52.39  & 4.01          & 145.53     & 11.15 \\
\hline
\end{tabular}
\end{table}


\subsection{$\sqrt{s}=7$ TeV case}
\label{subsec:7}
\begin{figure}[H]
\begin{center}
\centering
\resizebox{9.cm}{8.cm}{
\includegraphics[width=0.5\textwidth]{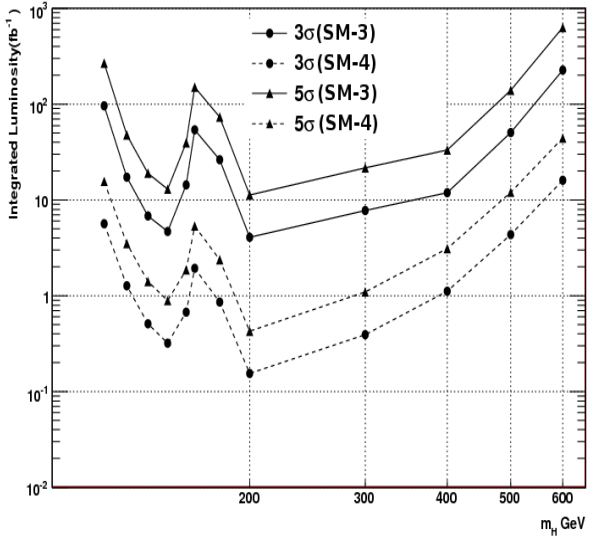}}
\caption{Integrated luminosity needed at 7 TeV for 3$\sigma$ and 5$\sigma$ for $gg \rightarrow H \rightarrow 
ZZ^{(*)}\rightarrow4\ell$ channel considering SM-3 and SM-4 cases.}
\label{fig:int7}
\end{center}
\end{figure}

Table \ref{tab:xscn7} shows expected cross sections after reconstruction of $gg\rightarrow H \rightarrow 
ZZ^{(*)}\rightarrow4\ell$ channel and its backgrounds. Corresponding statistical significance for 
various integrated luminosities are shown in Table \ref{tab:ss7} and $L_{int}$ needed for $3\sigma$ and $5\sigma$ 
significance are given in Table \ref{tab:int7} and plotted in Figure \ref{fig:int7}. 200 pb$^{-1}$ will scan 200 - 250 GeV Higgs 
whereas an additional 200 pb$^{-1}$ will scan up to 300 GeV. 

\begin{table}[H]
\caption{Expected cross sections of the reconstructed $gg\rightarrow H\rightarrow ZZ^{(*)}$$\rightarrow4\ell$
channel and its total background at 7 TeV.}
\label{tab:xscn7}
\tiny
\centering
\begin{tabular}{|c||c|c|c|}
\hline
         \multicolumn{4}{|c|}{ $gg\rightarrow H \rightarrow ZZ^{(*)}$$\rightarrow4\ell$ at 7 TeV}  \\
\hline
$ \rm {m_H}$ (GeV)& \multicolumn{3}{|c|}{ Cross section (fb) }           \\
\hline
                & \multicolumn{2}{|c|}{\bf Signal}                      & \multicolumn{1}{|c|}{\bf Background} \\
                & SM-3          & SM-4        & SM-3 \& SM-4   \\
\hline
 120            & 0.085         & 0.502      & 0.053  \\
 130            & 0.240         & 1.44       & 0.053  \\
 140            & 0.434         & 2.84       & 0.051  \\
 150            & 0.544         & 3.99       & 0.046  \\
 160            & 0.282         & 2.39       & 0.060  \\
 165            & 0.131         & 1.14       & 0.068  \\
 180            & 0.347         & 2.99       & 0.255  \\
 200            & 1.689         & 14.14      & 0.830  \\
 300            & 0.892         & 6.08       & 0.443  \\
 400            & 0.608         & 2.69       & 0.325  \\
 500            & 0.262         & 1.096      & 0.306  \\
 600            & 0.105         & 0.457      & 0.245  \\
\hline
\end{tabular}
\end{table}
\begin{table}[H]

\caption{Expected statistical significance for various integrated luminosity values at 7 TeV.}
\label{tab:ss7}

\tiny
\centering
\begin{tabular}{|c|c|c|c|c|c|c|c|c|}
\hline
         \multicolumn{9}{|c|}{ $gg\rightarrow H \rightarrow ZZ^{(*)}$$\rightarrow4\ell$ at 7 TeV }  \\
\hline
m$_H$   & \multicolumn{2}{|c|}{0.1fb$^{-1}$ } & \multicolumn{2}{|c|}{0.2fb$^{-1}$ } & 
\multicolumn{2}{|c|}{0.3fb$^{-1}$ }& \multicolumn{2}{|c|}{0.5fb$^{-1}$ } \\
                & SM-3 & SM-4           & SM-3 & SM-4           & SM-3 & SM-4           & SM-3 & SM-4 \\
\hline
 120              &0.097  & 0.400 			&0.137  &0.566 			&0.167  &0.693			&0.216   &0.894    \\
 130              &0.229  & 0.843			&0.323  &1.192 			&0.396  &1.460			&0.511   &1.885    \\
 140              &0.363  & 1.330			&0.514  &1.880			&0.629  &2.303			&0.812   &2.974    \\
 150              &0.438  & 1.677			&0.620  &2.371			&0.759  &2.904			&0.980   &3.749    \\
 160              &0.251  & 1.157			&0.355  &1.637			&0.435  &2.004			&0.561   &2.587    \\
 165              &0.129  & 0.684			&0.182  &0.967			&0.223  &1.184			&0.288   &1.529    \\
 180              &0.184  & 1.024			&0.261  &1.449			&0.320  &1.774			&0.413   &2.291    \\
 200              &0.471  & 2.416			&0.666  &3.417			&0.815  &4.184			&1.053   &5.402    \\
 300              &0.341  & 1.514			&0.482  &2.141			&0.590  &2.623			&0.762   &3.386    \\
 400             & 0.274 & 0.897			&0.388  &1.268			&0.475  &1.553			&0.614   &2.005    \\
 500              &0.134  & 0.455			&0.189  &0.644			&0.232  &0.789			&0.299   &1.019    \\
 600              &0.063  & 0237			&0.089  &0.335			&0.109  &0.411			&0.141   &0.530   \\
\hline
\end{tabular}
\end{table}
\begin{table}[H]
\caption{Integrated luminosity needed for 3$\sigma$ and 5$\sigma$ significance at 7 TeV.}
\label{tab:int7}
\tiny
\centering
\begin{tabular}{|c||c|c|c|c|}
\hline
\multicolumn{5}{|c|}{ $gg\rightarrow H \rightarrow ZZ^{(*)}$$\rightarrow4\ell$ at 7 TeV} \\
\hline
$ \rm {m_H}$ (GeV)      & \multicolumn{4}{|c|}{ Luminosity (fb$^{-1}$)} \\
\hline
                        & \multicolumn{2}{|c|}{\bf for 3$\sigma$}               & \multicolumn{2}{|c|}{\bf for 5$\sigma$} \\
                        & SM-3   & SM-4          & SM-3       & SM-4\\
\hline

120			&96.4   &5.62			&268   &15.6		\\
130			&17.2   &1.27			&47.8   &3.52		\\
140			&6.82   &0.51			&18.9   &1.41		\\
150			&4.68   &0.32			&13.0   &0.89		\\
160			&14.3   &0.67			&39.7   &1.87		\\
165			&54.0   &1.92			&150   &5.35		\\
180			&26.4   &0.86			&73.4   &2.38		\\
200			&4.06   &0.15			&11.3   &0.43		\\
300			&7.75   &0.39			&21.5   &1.09		\\
400			&11.9   &1.12			&33.2   &3.11		\\
500			&50.3   &4.34			&140   &12.1		\\
600			&227   &16.0			&632   &44.4		\\
\hline
\end{tabular}
\end{table}

\subsection{300 GeV Higgs with 500 GeV fourth family}
\label{subsec:special}
If the common Yukawa coupling constant is equal to SU(2) gauge coupling $g_w$ then flavour democracy 
predicts the mass value about 500 GeV for fourth family quarks. It is interesting to note that this mass value also allows to 
explain several "anomalies" in $B$, $B_s$ mixings and decays involving CP observables \cite{amarjit1,amarjit2}. If one considers the 
quartic coupling constant of the Higgs self interaction also to be equal to $g_w$, the Higgs boson mass is predicted to be 
around 300 GeV. In such a case the enhancement factor in $gg\rightarrow H$ production is 7 (Figure \ref{fig:enh3}) 
and the $H\rightarrow ZZ$ branching ratio is 0.3 (Figure \ref{fig:br4}). The integrated 
luminosity to achieve $3\sigma$ and $5\sigma$ significance in such a situation at different center of mass energies are shown 
in Table \ref{tab:h300}.

\begin{table}[H]
\caption{Integrated Luminosities (in fb$^{-1}$) needed to achieve 3 or 5 $\sigma$ significance at different 
center of mass energies.}
\label{tab:h300}       
\begin{center}
\begin{tabular}{|r|c|c|}
\hline
Energy&$L_{int}$(fb$^{-1}$) for $3\sigma$  	&$L_{int}$(fb$^{-1}$) for $5\sigma$  \\
\hline
14 TeV &0.07&0.21  \\
10 TeV &0.16&0.44  \\
7 TeV  &0.39&1.09  \\
\hline
\end{tabular}
\end{center}
\end{table}

\section{Conclusions}
\label{sec:conc}
Assuming that nature prefers the SM-4 case, Fermilab already excludes Higgs masses up to 200 GeV.  
In contrast to SM-3, in SM-4 case electroweak precision data favors a heavier Higgs \cite{kribs}. 
Hence, during the next couple of years we will experience tough competition between the two hadron 
colliders: running Tevatron and soon to run LHC. In our opinion, corresponding 
experiments at both machines should seriously consider SM-4 predictions.

\begin{acknowledgement}
We are grateful to V. N. $\c S$enoguz and G. \"{U}nel for useful discussion and crucial 
remarks. This work is 
supported in part by the Turkish Atomic Energy Authority (TAEK).
\end{acknowledgement}


\begin{thebibliography}{}

\bibitem{fritzsch}
H. Fritzsch, \textit{"Light neutrinos, nonuniversality of leptonic weak interaction and a fourth massive generation"}  
, MPI-PH-92-42, Phys. Lett. B \textbf{289} (1992) 92.

\bibitem{datta}
A. Datta, \textit{"Flavor democracy calls for the fourth generation"}, hep-ph/9207248, PRINT-92-0285 (JADAVPUR), 
IC-92-164, Pramana \textbf{40} (1993) L503.

\bibitem{celikel}
A. Celikel, A. K. Ciftci and S. Sultansoy, \textit{"A Search for fourth SM family"}, AU-HEP-94-12, 
Phys. Lett. B \textbf{342} (1995) 257.

\bibitem{sultansoy1}
S. Sultansoy, \textit{"Four ways to TeV scale"}, AU-HEP-97-04, Turk. J. Phys. \textbf{22} (1998) 575. 

\bibitem{sultansoy2}
S. Sultansoy, \textit{"Four remarks on physics at LHC"}, AU-HEP-97-05, Invited talk given at the ATLAS week, Geneva, Switzerland, 

\bibitem{frampton}
P. H. Frampton, P. Q. Hung and M. Sher, \textit{"Quarks and leptons beyond the third generation"}, hep-ph/9903387, IFP-759-UNCA, 
WM-99-104, Phys. Rept. \textbf{330} (2000) 263.

\bibitem{sultansoy3}
S. Sultansoy, \textit{"Why the four SM families"}, AU-HEP-00-04, hep-ph/0004271, (2000).

\bibitem{sultansoy4}
S. Sultansoy, \textit{"Flavor Democracy in Particle Physics"}, hep-ph/0610279, AIP Conf. Proc. \textbf{899} (2007) 49. 

\bibitem{sultansoy5}
S. Sultansoy, \textit{"The Naturalness of the Fourth SM Family"}, arXiv:0905.2874[hep-ph], (2009).

\bibitem{holdom}
B. Holdom \textit{et al}, \textit{"Four Statements about the Fourth Generation."}, arXiv:0904.4698[hep-ph], (2009).

\bibitem{he}
H.-J. He, N. Polonsky and S. Su, \textit{"Extra Families, Higgs Spectrum and Oblique Corrections"},   arXiv:hep-ph/0102144, 
CALT-68-2313, MIT--CTP--3080, UT-HEP-01-018, Phys. Rev. D \textbf{64} (2001) 053004.

\bibitem{novikov}
V. A. Novikov \textit{et al}, \textit{"Extra generations and discrepancies of electroweak precision 
data"}, arXiv:hep-ph/0111028, Phys. Lett. B \textbf{529} (2002) 111.

\bibitem{kribs}
G. D. Kribs \textit{et al}, \textit{"Four generations and Higgs physics"}, arXiv:0706.3718 [hep-ph], 
ANL-HEP-PR-07-39, Phys. Rev. D \textbf{76} (2007) 075016.

\bibitem{ozcan}
V. E. Ozcan, S. Sultansoy and G. Unel, \textit{"Possible Discovery Channel for New Charged Leptons at the LHC."}, arXiv:0903.3177, 
J. Phys. G \textbf{36} (2009) 095002.


\bibitem{b3sm}
"Beyond the 3SM generation at the LHC era" workshop, CERN, 04-05 September 2003 http://indico.cern.ch/conferenceDisplay.py?confId=33285

\bibitem{arik1}
E. Ar{\i}k \textit{et al}, \textit{"Enhancement of the Standard Model Higgs Boson Production Crosssection
with the Fourth Standard Model Family Quarks"}, ATLAS Internal Note ATL-PHYS-98-125 (1998).

\bibitem{atlas-tdr}
ATLAS Collaboration, \textit{"ATLAS TDR"}, \textbf{V2} Ch18, CERN/LHCC/99-15 (1999).

\bibitem{ginzburg}
I. F. Ginzburg, I. P. Ivanov and A. Schiller, \textit{"Search for fourth generation quarks and leptons at the Fermilab 
Tevatron and CERN Large Hadron Collider"}, hep-ph/9802364, UL-NTZ-03-98, Phys. Rev. D \textbf{60} (1999) 095001.

\bibitem{sultansoy6}
S. Sultansoy, \textit{"The 'Golden mode' at the upgraded Tevatron?"}, hep-ex/0010037, (2000).

\bibitem{arik2}
E. Ar{\i}k \textit{et al}, \textit{"With four standard model families, the LHC could discover the Higgs boson with a few 
fb$^{-1}$"}, hep-ph/0109037, ATLAS Scientific Note SN-ATLAS-2001-005, Eur. Phys. J. C \textbf{26} (2002) 9.

\bibitem{cakir1}
O. {\c C}ak{\i}r and S. Sultansoy, \textit{"The Fourth SM family enhancement to the golden mode at the upgraded Tevatron"}, 
hep-ph/0106312, Phys. Rev. D \textbf{65} (2002) 013009.

\bibitem{arik3}
E. Ar{\i}k \textit{et al}, \textit{"Consequences of the extra SM families on the Higgs boson production at Tevatron and CERN 
LHC"}, hep-ph/0203257, Phys. Rev. D \textbf{66} (2002) 033003.

\bibitem{arik4}
E. Ar{\i}k \textit{et al}, \textit{"Observability of the Higgs boson and extra 
SM families at the Tevatron"}, hep-ph/0502050, Acta Phys. Polon. B \textbf{37} (2006) 2839.


\bibitem{arik5}
E. Ar{\i}k, S. A. {\c C}etin and S. Sultansoy, \textit{"The impact of the fourth SM family on the Higgs 
observability at the LHC"}, arXiv:0708.0241 [hep-ph], Balk. Phys. Lett. \textbf{15} N4 (2007) 1.

\bibitem{barger}
V. D. Barger and R. J. Phillips, \textit{"Collider Physics"}, Addidon-Wesley (1987).

\bibitem{gokhan1}
S. Sultansoy and G. \"{U}nel, \textit{"'Silver' mode for the heavy Higgs search in the presence of a fourth SM family"}, 
arXiv:0707.3266 [hep-ph], Turk. J. Phys. \textbf{31} (2007) 295.

\bibitem{gokhan2}
T. {\c C}uhadar-Donszelmann \textit{et al}, \textit{"Fourth Family Neutrinos and the Higgs Boson"}, 
arXiv:0806.4003 [hep-ph], JHEP \textbf{0810} (2008) 074.

\bibitem{fnal-cdf}
CDF Collaboration, \textit{"CDF Search for Higgs to WW* Production using a Combined Matrix Element and Neural Network 
Technique"}, 
$http://www-cdf.fnal.gov/physics/new/hdg/results/hwwmenn_090710/$ (2009).

\bibitem{fnal-d0}
D0 Collaboration, \textit{"Search for Higgs Boson Production in Dilepton plus Missing Transverse Energy
Final States with 3.0{4.2 fb1 of pp Collisions at ps = 1:96 TeV}"}, D0 Note 5871-CONF (2009).

\bibitem{tevatron1}
W. M. Yao for the CDF and D0 Collaborations, \textit{"Neutral Higgs boson search at Tevatron"}, FERMILAB-CONF-04-307-E, 
Beijing ICHEP 2004, vol.2, 1228, hep-ex/0411053, (2004).

\bibitem{tevatron2}
V. Buescher, \textit{"Searches for Higgs bosons and supersymmetry at the Tevatron"}, FERMILAB-CONF-04-356-E, 
Tsukuba 2004, 'Supersymmetry and unification of fundamental interactions' 17-32, hep-ex/0411063, (2004).

\bibitem{tevatron3}
W. M. Yao, \textit{"Top quark and Higgs physics at the Tevatron"}, LBNL-56547, (2004)

\bibitem{tevatron4}
V. Buescher and K. Jakobs, \textit{"Higgs boson searches at hadron colliders"}, hep-ph/0504099, 
Int. J. Mod. Phys. A \textbf{20} (2005) 2523.

\bibitem{tevatron5}
V.M. Abazov \textit{et al}; by D0 Collaboration, \textit{"Search for the Higgs boson in H $\rightarrow$ WW$^{(*)}$ decays in p 
anti-p 
collisions at $\sqrt{s}$ = 1.96 TeV"}, FERMILAB-PUB-05-377-E, hep-ex/0508054, Phys. Rev. Lett. \textbf{96} (2006) 011801.

\bibitem{tevatron6}
A. Abulencia \textit{et al}; by CDF Collaboration , \textit{"Search for a neutral Higgs boson decaying to a W boson pair in p 
antip collisions at $\sqrt{s}$ = 1.96 TeV"}, FERMILAB-PUB-06-154-E, hep-ex/0605124, Phys. Rev. Lett. \textbf{97} 
(2006) 081802.

\bibitem{tevatron7}
A. Soha for the CDF Collaboration, \textit{"New phenomena searches at CDF"}, FERMILAB-CONF-06-157-E, (2006).

\bibitem{tevatron8}
A. Meyer, \textit{"Searches for New Phenomena at the Tevatron and at HERA"}, FERMILAB-CONF-06-404-E, hep-ex/0610001, (2006).

\bibitem{tevatron9}
G. Bernardi for the D0 Collaboration, \textit{"Searches and Prospects for the Standard Model Higgs boson at the Tevatron"}, 
ICHEP 06, hep-ex/0612046, (2006).

\bibitem{tevatron10}
T. Nunnemann for the D0 and CDF Collaborations, \textit{"Searches for the Higgs boson and supersymmetry at the Tevatron"}, 
arXiv:0710.0248 [hep-ex], Frascati Phys. Ser. \textbf{44} (2007) 475.

\bibitem{atlas-jinst}
ATLAS Collaboration, \textit{"The ATLAS Experiment at the CERN Large Hadron Collider"}, Journal of Instrumentation \textbf{3} (1999) 
S08003.

\bibitem{atlas-csc}
ATLAS Collaboration, \textit{"Expected Performance of the ATLAS Experiment - Detector, Trigger and Physics"}, arXiv:0901.0512 
[hep-ex] (2009).

\bibitem{higlu}
M. Spira, \textit{"HIGLU"}, DESY-T-95-05, hep-ph/9510347 (1995).

\bibitem{comphep}
CompHEP Collaboration, \textit{"Automatic computations from Lagrangians to events"}, arXiv:hep-ph/0403113, Nucl. Instrum. Meth. 
A\textbf{534} (2004) 250;
A.Pukhov \textit{et al}, \textit{"CompHEP - a package for evaluation of Feynman diagrams and integration over multi-particle phase 
space}, INP MSU report 98-41/542, arXiv:hep-ph/9908288 (1998).

\bibitem{bruce}
J. M. Campbell \textit{et al}, \textit{"Normalizing Weak Boson Pair Production at the Large Hadron Collider"},    
arXiv:0906.2500 (2009).

\bibitem{amarjit1}
A. Soni \textit{et al}, \textit{"The Fourth family: A Natural explanation for the observed pattern of anomalies in B-CP 
asymmetries."}, arXiv:0807.1971 (2008).

\bibitem{amarjit2}
A. Soni, \textit{"The '4th generation', B-CP anomalies and the LHC."}, arXiv:0907.2057 (2009).



\end{thebibliography}

\end{document}